# A Physics-Informed Bayesian Optimization Method for Rapid Development of Electrical Machines

Pedram Asef, Department of Mechanical Engineering, University College London, London, E20 3BS, UK

Christopher Vagg, Department of Mechanical Engineering, University of Bath, Bath, BA2 7AY, UK

*Abstract*— **Advanced slot and winding designs are imperative to create future high performance electrical machines (EM). As a result, the development of methods to design and improve slot filling factor (SFF) has attracted considerable research. Recent developments in manufacturing processes, such as additive manufacturing and alternative materials, has also highlighted a need for novel high-fidelity design techniques to develop high performance complex geometries and topologies. This study therefore introduces a novel physics-informed machine learning (PIML) design optimization process for improving SFF in traction electrical machines used in electric vehicles. A maximum entropy sampling algorithm (MESA) is used to seed a physics-informed Bayesian optimization (PIBO) algorithm, where the target function and its approximations are produced by Gaussian processes (GP)s. The proposed PIBO-MESA is coupled with a 2D finite element model (FEM) to perform a GP-based surrogate and provide the first demonstration of the optimal combination of complex design variables for an electrical machine. Significant computational gains were achieved using the new PIBO-MESA approach, which is 45% faster than existing stochastic methods, such as the non-dominated sorting genetic algorithm II (NSGA-II). The FEM results confirm that the new design optimization process and keystone shaped wires lead to a higher SFF (i.e. by 20%) and electromagnetic improvements (e.g. maximum torque by 12%) with similar resistivity. The newly developed PIBO-MESA design optimization process therefore presents significant benefits in the design of high-performance electric machines, with reduced development time and costs.**

## I. Introduction

*A. Background*:

Among all types of synchronous motors without electric current in their rotor, permanent magnet synchronous machines (PMSM)s gained significant consideration in the electric vehicle (EV) market recently [1-2]. This is due to their high performance, such as high torque at low operational speeds and efficient and reliable inverters. However, they are still far from ideal because of performance and manufacturing issues. From a performance perspective, there are risks associated with: (i) large braking torque when the traction inverter fails; (ii) maximum speed limitations caused by permanent magnets (PM) which can induce high voltages in stator windings; and (iii) no control over the current (i.e. power) in the rotor. From a manufacturing perspective, (i) the usage of rare-earth (RE)-based PMs is critical due to their environmental and sustainability concerns. Thereby enormous research interests are focused on PM-free motor topologies; (ii) expensive; and (iii) mechanically weak (e.g. breakable and damageable due to temperature rise and demagnetization). Researchers are nowadays working on many alternative alternating current (AC) motors which do not use PMs (e.g. reluctance motor (RM)) [3] or hybrid options with a lesser volume of PMs (e.g. PM-assisted switched reluctance motor [4]). So far, these are not competitive with PMSMs in electromagnetic performance [5-7]. A promising magnet-free and brushless design topology is AC electrically excited synchronous motors (EESM)s with an inductive power transfer system as presented in Fig. 1. The power transfer system consists of rotary and static coils to inject controllable direct current (DC) in the rotor windings. The EESMs were not being widely investigated yet. After the electrically switching excitation circuit invention [8], the EESM is revived as a promising candidate for further development. Limited works highlight their capabilities [9-12] for traction applications. The prime benefits of this topology are: (i) cost-effective, the total elimination of magnets which makes EESM much cheaper than other topologies using magnets (e.g. PMSMs), (ii) a more sustainable and environmentally friendly choice due to extremely high demands of PM materials and market limitations; (iii) excellent power factor which allows maximum torque improvement; (iv) increasing magnetic flux density by injecting more current in the rotor winding. The control over the rotor's current and magnetic field helps with maximum torque density maximization; (v) increasing efficiency by minimizing electromagnetic loss, mainly during low torque operation; and (vii) safer operation, as the magnetic field in the rotor is controllable, allowing fast demagnetization to overcome hazardous braking torques. The brushed separately excited synchronous motor (SSM) topology was discovered more than

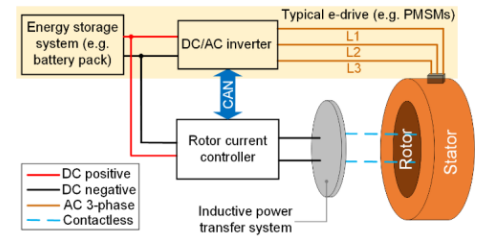

Fig. 1. Traction electric drive system and additional components for EESMs.

Pedram Asef is with Advanced Propulsion Laboratory (APL), Department of Mechanical Engineering, University College London, E20 3BS, UK.

Christopher Vagg is with Institute for Advanced Automotive Propulsion Systems (IAAPS), Faculty of Engineering & Design, University of Bath, BA2 7AY, Bath, Somerset, UK. *(Corresponding author: P.Asef). Email:*pedram.asef@ucl.ac.uk*; Tel.:* +44 20 7504 5287.

a century ago, however they were not chosen for automotive applications because of their lower performance, e.g., torque production challenges. To date, many researchers are working on SSMs without brushes and slip rings for injecting electric current into the field (or rotor) windings. This problem was resolved by the researchers, more details are given in [8]. They invented a switching exciter circuit in the side of the rotor winding to transfer current from the power supply to the rotor of the EESMs. A few researchers investigated the analytical modelling of EESMs [9], in which they studied a formalized method to calculate the field distribution and *dq*-axis model parameters. In another work [10], the researchers proposed a low-order lumped parameter thermal network to predict the temperature rise of the EESMs, in which thermal resistances and capacities were identified individually. C. Stancu *et al.* [11], studied a contactless rotary transformer and field converter for the excitation of the rotor winding. The researchers found that the EESM is capable of considerably more peak torque and power compared to high-performing interior permanent magnet synchronous motors (IPMSM)s. They were thermally limited by the rotor coils to approximately 30s of performance at that level, but continuous power and torque also exceeded that of IPMSM. Maximum motor efficiency higher than 95%, including the losses of the motor, the rotating transformer, and the field converter was achieved. Overall, excellent efficiencies at high speeds were recorded, outperforming an equivalent optimized IPMSM with a similar size. At lower speeds, the IPMSM was more efficient. G. Jawad *et al.* [12], introduced a new approach to excite the rotor winding of EESM. They utilized zero-sequence third-harmonic currents to generate an additional third-harmonic magneto-motive force (MMF) pulsating spatially in the stator winding. The proposed topology can be employed either as an inverter or a conventional AC power source to generate the harmonic component of the air-gap flux for rotor excitation and the fundamental component to produce the torque. In [13], the authors presented a comprehensive review on low-cost high-performance ferrite magnets which are more sustainable and have a more stable supply chain for EV applications.

In summary, Table 1 compares the most attractive traction motors qualitatively. Among these, the EESM is the most favourable and researchable topology for future EVs. Because of its inexpensive total material cost and unique power factor (close to one), indicating that most of the inverter power can be transferred to mechanical power at the shaft. The PMSM's unique feature is performance, high torque at low speeds and premium efficiency at low and high speeds, due to its high power density capability [5-7]. However, they are significantly more expensive than other topologies because of PMs, in addition to their environmental and sustainability concerns. The low safety of PMSMs is mainly due to critical braking torque production if the inverter does not function properly (e.g. short-circuited). The induction motor (IM), without the usage of PMs and any mechanical contact with the moving parts, can be considered a safe and cheap topology. They suffer to provide a constant torque below the nominal speeds, in other words, the IMs fail to offer a constant maximum power over a wide speed range. Hence, they require a variable-speed transmission gearbox for EV applications [14]. The RMs are the cheapest topology, and yet unpopular because of their poor performance e.g., their maximum and continuous power capability is the lowest and torque density is also very low.

TABLE I QUALITATIVE COMPARISON OF ELECTRICAL MOTORS

| Factor/ type | EESM | PMSM | IM | RM |
|---|---|---|---|---|
| Power density | M | H* | M | L |
| Power factor | H* | H | M | L |
| Torque density | H | H* | H | L |
| CPSR capability | H | H | M | L |
| Efficiency | M | H | H | M |
| Material cost | L | H | L* | L* |
| Environmentally friendly | H | L | H | H |
| Maintenance | M | M | M | L |
| Safety | H | L | H | H |

**Note**: H, M, and L are representing high, medium, and low; * shows the unique feature of each traction electric machine topology.

*B. Related Work*:

The use of design optimization in electrical machines accomplished significant performance improvements and developments. Many researchers used deterministic methods [2][5][15-18], stochastic [19] (e.g. genetic [20-22] and particle swarm [23-25] algorithms) methods, and machine learning (ML) methods [1] [26-33], such as Bayesian optimization (BO) algorithms [31] and [34], to improve different aspects of motor performance. Most probabilistic ML methods offer a promising framework for understanding the uncertainty of the design problem and its performance predictions. Based on [35-36], well-regulated BO is essential to gain the best possible outcomes. To find the best Surrogate model for the BO, a comparative study [37] is conducted using the most popular surrogate models for calibration reasons, such as random forests (RF) [38], deep ensembles (DE) [39], Bayesian neural network (BNN) [40], and Gaussian processes (GP) [41]. It is found that GPs can work well with BO-based design optimizations. Applied ML methods are recently utilized to improve different design aspects of electric machines for demagnetization faults [42], bearing faults [43], short circuit faults [44], and control [45]. For example, Song *et al.*[42] investigated a new meta-learning method under varying operating conditions.

*C. Contribution*:

The research hypothesis is: Can a magnet-free electric motor provide similar performance to existing PM traction motors? This is under investigation by many researchers today. In this study, a recently invented EESM is electromagnetically investigated as a benchmark motor. A new physics-informed BO (PIBO) design optimization algorithm is developed to improve the slot fill factor (SFF) for the EESMs. The developed PIBO method dealt with the copper filling of stator trapezoidal slots using innovative keystone shaped Litz wires type 8 with as many as possible single insulated strands twisted and compressed into custom conductive profiles. The developed PIBO is driven by ML models, such as GP. Exclusively, a maximum entropy sampling algorithm (MESA) is also introduced for seeding the PIBO-MESA algorithm. The proposed PIBO-MESA could outperform some popular design optimization in terms of computation time by about 45%. The outcome of the proposed PIBO-MESA algorithm and use of keystone wires offer a 20% improvement in the SFF which also resulted in better electromagnetic performance. The developed PIBO-MESA can be used for different electric motors with trapezoidal slots used in their stators.

## II. Mathematical Modelling and Specification

The EESM topology creates more design complications for the motor compared to other popular traction motors, such as PMSMs and IMs, because of the current injection in the rotor winding by a transformer, which consists of static and rotating parts. In the following sections, the main mathematical equations and design specifications of the studied EESM are given.

### A. Mathematical Formulations

The impact of various rotor and stator coil turns is investigated to increase the SFF. The motivation is finding out the optimum design choices for desired inductance and torque improvement. Considering the electromagnetic torque as:

$$T = \frac{3}{2} I \Psi \cos(\varphi) \quad (1)$$

where maximizing the torque density simply depends on magnetic flux $\Psi$ and power factor $\cos(\varphi)$ for a supplied current $I$. Therefore, the torque can be increased if: (1) the resistance of stator and rotor windings are reduced with consideration of the same current baseline, and/or (2) the power factor is improved. The supplied currents based on magnetic flux, power factor, and torque can be calculated as a function of speed:

$$\Psi = \begin{pmatrix} \Psi_d \\ \Psi_q \end{pmatrix} = \begin{pmatrix} L_d \cdot I_d + L'_r \cdot I_r \\ L_q \cdot I_q \end{pmatrix} \quad (2)$$

where $dq$-axis inductances ($L_d$ and $L_q$) are a function of $dq$-axis currents ($I_d$ and $I_q$) as well as rotor current $I_r$. The rotor inductance $L'_r$ multiplied by $I_r$ provides stator flux.

The total (stator and rotor) conduction losses are given as:

$$P_{Jtot} = \frac{3}{2} R_s I^2 + R_r I_r^2 \quad (3)$$

The iron loss, based on [15], with consideration of hysteresis and eddy current losses, is calculated using:

$$P_{fetot} = \Sigma \left( \left( k_h \left( \frac{f}{f_0} \right)^\alpha + k_e \left( \frac{f}{f_0} \right)^\beta \right) \left( \frac{B}{B_0} \right)^\gamma \rho V \right) k_u \quad (4)$$

where $k_h$ is the hysteresis coefficient at $f_0$ and $B_0$ in W/kg, which depends on the core material type and thickness, $k_e$ is the eddy current coefficient at $f_0$ and $B_0$ in W/kg, mainly restricted by the lamination thickness, $\alpha$, $\beta$, and $\gamma$ are coefficients (in range of 1-2) that depend on the properties of the iron core material, $f$ and $f_0$ represent electrical and reference frequencies, $B$ and $B_0$ are magnetic flux density and reference magnetic flux density, computed using a 2D FEA solver, $\rho$ is volumetric mass density, $V$ is an element volume, and $k_u$ is field factor that has a multiplicative effect.

Taking the motor pole number ($p = 6$) and seven windings in each slot being connected in parallel, this indicates that two windings are in each slot if they are connected in series. Considering the benefits of trapezoidal slot shapes to reduce the winding resistance and leakage inductance, the optimum size of each Litz wires bar is computed for the stator slot:

$$I_{e,t} = \frac{\pi \cdot f \left( h/N_c \right)^2 B \cdot W \cdot \sigma_c}{4} \quad (5)$$

where $h$ is the bar height, $N_c$ is the number of conductors, $B$ is the magnetic flux density, $W$ is the bat's width, $\sigma_c$ is the conductivity of copper. As $B$ is almost in a tangential direction, the use of fixed dimension bars (consisting of several conductors) is unsuitable. Additionally, the use of single bars has manufacturing challenges because of parallel connection requirements. Hence, the four identical, type eight Litz wires topology was selected as the benchmark model presented in Fig. 2. This Litz wire type contains single insulated magnetic wire strands twisted and flattened into a rectangular shape with optional outer insulation of textile yarn. This topology allows customizing the bar sizes to fill in the slots thoroughly, which offers excellent SFF [15-17].

To achieve superior electromagnetic performance (e.g. efficiency, peak torque, and continuous power), the maximization of SFF in the stator and rotor is vital. Particularly, the stator Joule loss (first term in Eq. 3) plays a considerable role in power losses at a wide range of speeds, and the stator resistance has an inverse proportional relationship to SFF. In other words, increased SFF leads to lower stator Joule loss, suppressing the motor's temperature rise. For this reason, a new design optimization process is introduced, in section B, to further increase SFF and the EESM's electromagnetic means.

### B. Design Specification of Reference EESM

The reference motor design specification is inspired by [8]; the stator and rotor are made of M270-35A steel laminations with 0.95 stacking factor, dimensions are given in Table II. The EESM uses distributed winding topology with two layers, in which the windings are limited to two bars inserted in each slot, each bar consists of 65 copper wires with a diameter of 0.8 mm for the stator core. The studied motor topology can offer excellent power over a wide range of speeds compared to other attractive traction motors as demonstrated in Fig. 2.

TABLE II Main Design Parameters for Reference EESM

| P. | Description | Value | Unit |
|---|---|---|---|
| $R_{os}$ | Outer stator radius | 120 | mm |
| $R_{is}$ | Inner stator radius | 83.0 | mm |
| $R_{or}$ | Outer rotor diameter | 81.5 | mm |
| $R_{ir}$ | Inner rotor diameter | 40.0 | mm |
| $l_g$ | Maximum airgap length | 1.5 | mm |
| $l_c$ | Axial length of the motor | 110 | mm |
| SPP | Slot per pole per phase | 2.0 | - |
| SFF | Slot filling factor | 60.0 | % |
| SF | Stack factor | 95.0 | % |
| WpS | Number of windings per slot | 7.0 | - |
| SPW | Number of stator parallel windings | 3.0 | - |
| $I_{max}$ | Maximum inverter current | 635.0 | A |
| $V_{dc,link}$ | DC-link (battery) voltage | 400.0 | V |

### C. Novel Design Optimization Process and Algorithms

To maximize the SFF of the EESM, several related changeable variables are chosen, as given in Table II. A fast and high-fidelity (PIBO-MESA) optimization algorithm is employed to tailor the conductive bars with four identical, type eight Litz wires. In this design process technique, a randomized maximum entropy sampling algorithm (MESA) is employed to fill the design space of each design variable [46-47] illustrated in Table III and Fig. 2. The min/max boundaries of each variable are restricted by the tooth width and yoke height for avoiding saturation and overheating in the motor. The MESA techniques are also used in other applications [18-19]. For different engineering problems, they can be tied with other local and global searchers to find global optimum solutions [20-23]. In this work, the supervised Gaussian-based model is uniquely tied to the MESA using a machine learning algorithm to find the best design selection for the copper in the stator slots of the EESM. In this design problem, the stator trapezoidal slots are assumed to host two bars, thereby the challenge is to find the optimum number of conductors to fit both bars within the slot dimensions considering the current density and temperature rise distributions, as presented in Fig. 2.

In Algorithm 1, the utilization of MESA introduces a new random approach to generate iterative samples for the PIBO in the search space. The objective of MESA is to select a most informative subset of $s$ random design variables from a set of $n$ random design variables, subject to side and/or logical constraints. The randomized variables are mostly assumed to be Gaussian, or that they can be properly transformed in many states. Assuming an optimal solution of $\hat{x} \in (0, \infty)$, sampling of a $s$-sized subset $S \subseteq [n]$ is generated with a probability of:

$$\mathbb{P}[\tilde{S} = S] = \frac{\prod_{i \in S} \hat{x}_i}{\sum_{S \in \binom{[n]}{s}} \prod_{i \in \bar{S}} \hat{x}_i} \quad (6)$$

A GP-based PIBO-MESA optimization algorithm is a probabilistic approach which utilizes a procedure to iteratively fit the probabilistic surrogate model to measured values of an objective function. The use of BO is effective when the objective function is expensive [35-37]. In this work, the MESA sampling mechanism is performed using Algorithm 1 to improve the random input data distributions for a high-fidelity design optimization process, particularly in initialization practice. The implementation and assumptions of the design optimization process in Algorithm 2 are explained next.

In the initialization phase, the samples generated by MESA using Algorithm 1 in the search domains are simulated. Next, based on [35], most GPs are initialized by uniform random methods using an instantaneous regret term $r_t$ to be minimized, see (7), a $\frac{\Lambda_0}{2}$ term is used for multi-level fidelity design optimization processes. In this study, two-level fidelity is considered; in which the first-level process used $\frac{\Lambda_0}{2}$ and the second-level process utilized $\frac{\Lambda_0}{2}$.

$$S(\Lambda) = \min_{t=1,\ldots,N} r_t = \begin{cases} \min_{\substack{t=1,\ldots,N \\ t:m_t=M}} f_{op} - f^M(x_t) & \text{if } M^{th} \text{ fidelity achieved} \\ +\infty & \text{otherwise} \end{cases} \quad (7)$$

where $m_{t\geq 0}$ is requested at every iteration, $N$ is the random quantity within all fidelities up to $\Lambda$. In this optimization, only when $f = f^M$ is queried; an instantaneous reward term $y_t$ is set $-\infty$ if $m_t \neq M$ and $f^M(x_t)$ if $m_t = M$. Homogenously, $r_t = f_{op} - y_t$ indicates $r_t = +\infty$, when $m_t \neq M$ and $f_{op} - f^M(x_t)$ if $m_t = M$. The minimum regret $S(\Lambda)$ could be achieved using the MESA.

In the experimentation phase, based one [35-37], the squared exponential (SE) type of kernel is employed. The SE kernel is initialized via maximizing the GPs marginal likelihood on the early sample filling in the SE kernel at every 20 iterations applying marginal likelihood.

The objective was to maximize the SFF subjected to several constraints, such as frequency, skin effect, current density, temperature, and saturation. More details are given in Algorithm 2. For each stator slot, the SFF is improved when the cost function is maximized using the proposed design optimization method, where the cost is:

$$f(x) = \frac{\sum_{j=1}^{j=n_{max}}(x_7(i,j).x_8(i,j))}{((x_1(i)+x_2(i))x_4(i).0.5} \quad (8)$$

where $j$ and $n$ are the number of bars and the maximum number of Litz wire bars in each stator slot, $i$ is the number of conductors per bar, and $nn$ is the maximum number of conductors. In this study, $n$ is the total number of bars up to maximum number of $n$ bars ($n_{max} = 200$). Of course, the maximization of (6) is restricted to (i) design variable search spaces, and (ii) equal and non-equal performance-related constraints given in Algorithm 2. In this algorithm, a high-fidelity working mechanism of PIBO working with GP is presented. A squared exponential (SE) kernel is employed as they are well-known kernels to work with GP algorithms. Because of their universal nature and compatibility to be integrated with most types of functions. Each function in its prior has many derivatives and two main parameters to compute: (1) the length scale $\ell$ that defines the length of the fluctuations, in the function, extrapolated within the threshold $\ell$ of the model. (2) the output variance $\sigma^2$ (or scale factor) which decides the average distance of the function away from its mean value. The seeding samples in the four-dimensional (4D) space are generated by Algorithm 1, to be searched iteratively and globally. The global search aimed at maximizing the cost-function (line 8) subjected to the constraint functions $g$ given in lines 9 to 14. $g_1^{(m_t,i,j)}(x_{t,i,j})$ is required cross-sectional area for the maximum current density $J_s$, where $a$ is the number of parallel pathways in the winding, $ph$ is the number of phases in the stator winding, $V$ is terminal voltage, $\eta$ indicates efficiency,

TABLE III MAIN DESIGN OPTIMIZATION PARAMETERS FOR EESM

| P. | Description | Ini. | Min | Max |
|---|---|---|---|---|
| $x_1$ | Top base of isosceles trapezoids ($b_1$) | 7.0 | 6.0 | 8.0 |
| $x_2$ | Low base of isosceles trapezoids ($b_2$) | 4.0 | 3.0 | 6.0 |
| $x_3$ | Leg height of isosceles trapezoids ($h_1$) | 21.6 | 19.6 | 23.6 |
| $x_4$ | Height of slot conductive part ($h_c$) | 20.9 | 18.9 | 22.9 |
| $x_5$ | First & second congruent angles (α, β) | 85.9 | 84.9 | 86.9 |
| $x_6$ | Individual conductor bar length ($C_l$) | 125 | 115 | 135 |
| $x_7$ | Individual conductor bar height ($C_h$) | 94.1 | 93.1 | 95.1 |
| $x_8$ | Individual conductor bar width ($C_w$) | 94.1 | 93.1 | 95.1 |
| $x_9$ | Slot bottom fillet radius ($S_{FR}$) | 5.0 | 3.0 | 8.0 |
| $x_{10}$ | Slot opening width ($W_o$) | 10.0 | 8.0 | 12.0 |

**Note** P. shows parameters and Ini. stands for initial values; all lengths are in mm and angles in degree (°).

**Algorithm 1**: MESA for Given Design Variables

**1:** Range of each input $x$: i.e. $x_i = (x_{i_{min}}, x_{i_{max}}) \in \mathbb{R}$

**2:** Variables space: $\wp = x_1 \times x_2 \times x_3 \ldots x_{12} \subset \mathbb{R}^3$

**3: Create** $n \times n$ matrix of $g \geq 0$ with rank of $d$ and integer $s \in [d]$

**4:** Let $\hat{x}$ be an optimum solution of:

$$z := \max_x \{\Gamma_s(\sum_{i \in [n]} x_i v_i v_i^T) : \sum_{i \in [n]} x_i = s, x \in [0,1]^n\}$$

**5: Initialize** a selected set of $S = \emptyset$ and rejected set of $T = \emptyset$

**6: Set** two iterative parameters $D_1 = \sum_{S \in \binom{[n]}{s}} \prod_{i \in S} \hat{x}_i$, and $D_2 = 0$

**7: for** $j = 1, \cdots, n$ **do**

**8:** Compute $D_2 = \sum_{\substack{S \in \binom{[n]\setminus S \cup T}{s-1-|S|}}} \prod_{\tau \in S} \hat{x}_\tau$

**9:** To sample a $(0,1)$ uniform random parameter $U$

**10:** **if** $\hat{x}_j D_2 / D_1 \geq U$ **then**

**11:** Insert $j$ to set $\tilde{S}$

**12:** $D_1 = D_2$, **else**

**13:** Insert $j$ to set $T$

**14:** $D_1 = D_1 - \hat{x}_j D_2$

**15: end if**

**16: end for**

**17: Store** outputs in $\tilde{S}$

**Algorithm 2**: High-fidelity PIBO-GP algorithm for slot filling

**1: Initialization:** $X_i^* = [x_1^*, x_2^*, \ldots, x_D^*]^\top \in \mathbb{R}$; WL = 2, 4, 6, 8, ..., 10

**2: Define** bounds $\{\zeta^{(m)}\}_{m=1}^M$ and thresholds $\{\gamma^{(m)}\}_{m=1}^M$ in SE kernel $k$

**3:** Create inputs $m = 1, \ldots, M$, in which:

$D_0^{(m)} \leftarrow \emptyset$ and $(\mu_0^{(m)}, \sigma_0^{(m)}) \leftarrow (0, k^{0.5})$

**4:** for $t \in \mathbb{Z}_{>0} = \{1,2,3,\ldots,T\}$, $j = 1,\ldots,n$, $i = 10,\ldots,nn$

**5:** $x_{t,i,j} \leftarrow \arg\max_{x \in \chi} \varphi_{t,i,j}(x) = \min_{m=1,\ldots,M} \mu_{t-1}^{(m)}(x) + \beta_t^{0.5} \sigma_{t-1}^{(m)}(x) + \varsigma^m$

**6:** $m_{t,i,j} = \min(m | \beta_{t,i,j}^{0.5} \sigma_{t-1}^{(m)}(x_{t,i,j}) \geq \gamma^m \text{ or } m = M)$

**7:** $y_{t,i,j} \leftarrow$ queries of $f^{(m_t,i,j)}(x_{t,i,j})$

**8: Set** $f^{(m_t,i,j)}(x_{t,i,j}) = \arg\max \frac{\sum_{j=1}^{j=n} \max(x_7(i,j).x_8(i,j))}{((x_1(i)+x_2(i))x_4(i).0.5} \in \mathbb{R}$

**9:** s.t. $g_1^{(m_t,i,j)}(x_{t,i,j}) = \frac{\frac{P_{mech.}(x_{t,i,j})}{ph.V.\eta.cos\theta}}{a.J_s(x_{t,i,j})}$

**10:** $g_2^{(m_t,i,j)}(x_{t,i,j}) = \hat{B}_{t,i,j} - \frac{A_s(x_{t,i,j})}{A_{s,fe}(x_{t,i,j})} \mu_o \hat{H}_{t,i,j}$

**11:** $g_3^{(m_t,i,j)}(x_{t,i,j}) = \frac{1.24\left(\frac{\hat{U}_s(x_{t,i,j})+\hat{U}_r(x_{t,i,j})}{\hat{U}_g(x_{t,i,j})}\right)+1}{1.42\left(\frac{\hat{U}_s(x_{t,i,j})+\hat{U}_r(x_{t,i,j})}{\hat{U}_g(x_{t,i,j})}\right)+1.6}$

**12:** $g_4^{(m_t,i,j)}(x_{t,i,j}) = \frac{2\pi.f.\hat{B}_{t,i,j}.x_6.x_7}{2}$

**13:** $g_5^{(m_t,i,j)}(x_{t,i,j}) = \frac{\pi.f.x_7^2.\hat{B}_{t,i,j}.x_8.\sigma_c}{4}$

**14:** $g_6^{(m_t,i,j)}(x_{t,i,j}) = T_{sw}@6k \& 12k \ rpm \geq 120 \& 160°C$

**15:** To **update:** $D_t^{(m_t)} \leftarrow D_{t-1}^{(m_t)} \cup \{(x_t, y_t)\}$

**16: Compute** $\mu_t^{(m_t)} = k^\top (K + \eta^2 I_t)^{-1} Y$

**17:** and $\sigma_t^{(m_t)} = k(x,x) - k^\top (K + \eta^2 I_t)^{-1} k$

**18:** where $D_t^{(m)} \leftarrow D_{t-1}^{(m)}; \mu_t^{(m)} \leftarrow \mu_{t-1}^{(m)}; \sigma_t^{(m)} \leftarrow \sigma_{t-1}^{(m)}$ when $m \neq m_t$

**19:** end for

**20: Store** $X_T^* = \arg\max_{x_{t,i,j}} \{y(x_{t,i,j})\}_{t=1}^T$

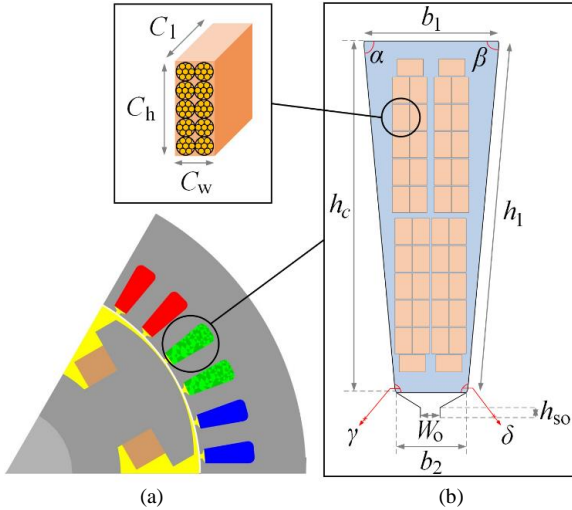

Fig. 2. Design optimization parameters, (a) the EESM 2-D model, and (b) the stator slot and four identical, type eight concentric Litz wires topology bars.

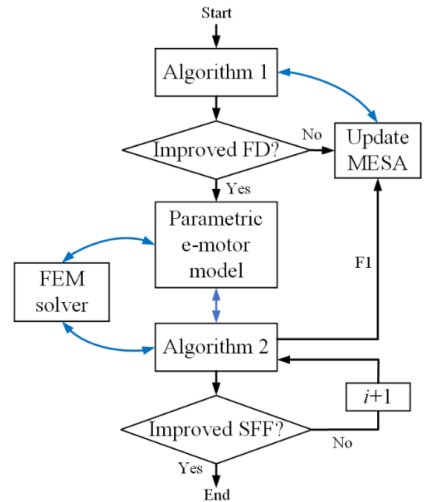

Fig. 3. The working mechanism of the proposed PIBO-MSEA design optimization process.

and $cos\theta$ is the power factor. $g_2^{(m_t,i,j)}(x_{t,i,j})$ considers the saturation by computing the fluxleakage corresponding with the field weakening in the stator tooth, where $\hat{H}_{t,i,j}$ is the field intensity in the iron tooth, $A_s(x_{t,i,j})$ is the slot area without the iron part, $A_{s,fe}(x_{t,i,j})$ is the iron part of the stator tooth region. $g_3^{(m_t,i,j)}(x_{t,i,j})$ computes the saturation factor, where $\hat{U}_s(x_{t,i,j})$, $\hat{U}_r(x_{t,i,j})$, and $\hat{U}_g(x_{t,i,j})$ are the magnetic potentials in the stator, rotor, and airgap regions in the EESM, respectively. $g_4^{(m_t,i,j)}(x_{t,i,j})$ takes into consideration the induced voltage in the conductors.

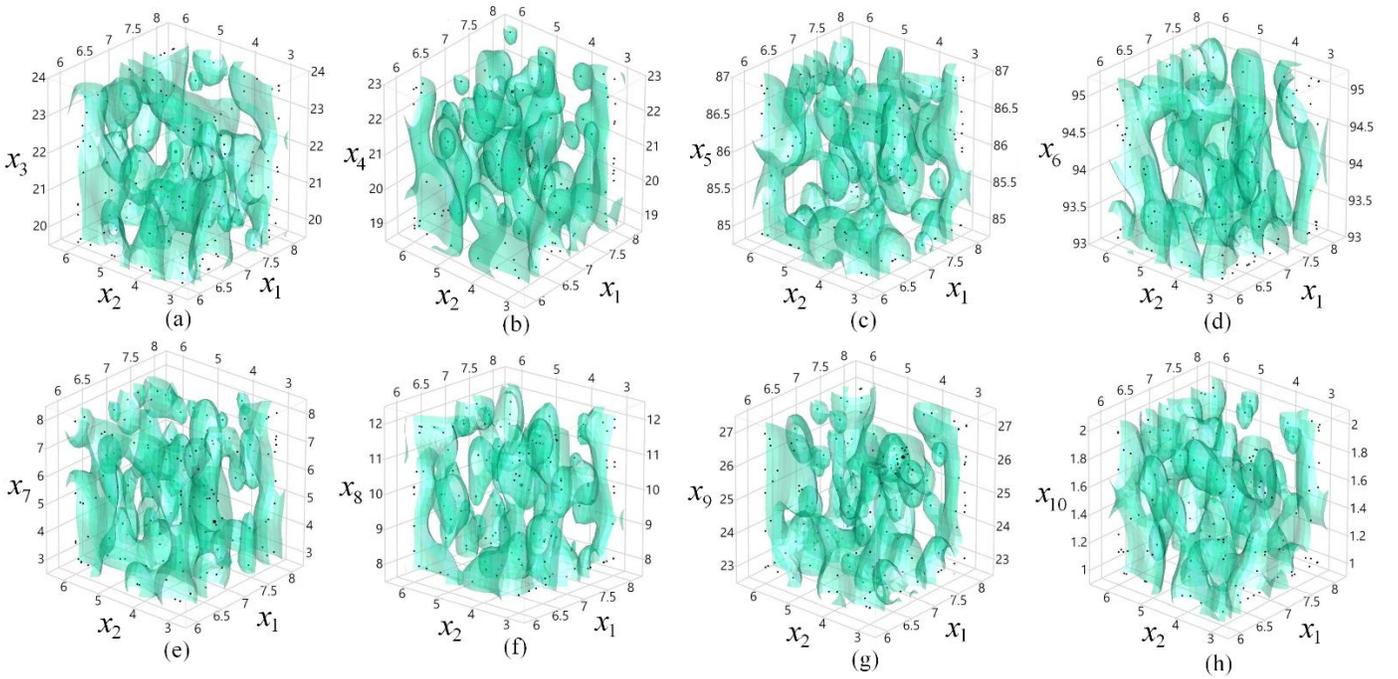

Fig. 4. 4D improved design optimization sample generation using PIBO-MESA (Algorithms 1 and 2) considering sample's mass density in the permitted design spaces for (a) $x_1,x_2,$ and $x_3$, (b) $x_1,x_2,$ and $x_4$, (c) $x_1,x_2,$ and $x_5$, (d) $x_1,x_2,$ and $x_6$, (e) $x_1,x_2,$ and $x_7$, (f) $x_1,x_2,$ and $x_8$, (g) $x_1,x_2,$ and $x_9$, and (h) $x_1,x_2,$ and $x_{10}$.

$g_5^{(m_t,i,j)}(x_{t,i,j})$ calculates the parasitic induced current induced in the conductor based on the main Eddy current loop, in which the current passes across the conductor. The parasitic induced current flows by delay of 90° compared to the current passing 160 A at 12000 rpm. All queries with $x$ and $y$ values updated the matrix $D_t^{(m_t)}$, among which the best $x$ combinations are stored in $\mathrm{X}_T^*$.

The working mechanism of the PIBO with the MESA is presented in Fig. 3, in which the start signal introduces the design variables available to influence the cost-function given in Algorithm 2. In the first part of the flowchart, the filling design points are distributed within the search bounds to fill the 4D cubes as cost-effectively as possible, as shown in Fig. 4. As the aim is to reduce the computational time, with assistance of Algorithm 2 (i.e. F1), the sampling distribution shrinks iteratively toward the global optimum region. Faster design optimization is achieved because both Algorithms 1 and 2 are physics-driven and well-calibrated. At the heart of the design optimization process, a 2D FEM solver is employed to simultaneously compute the EESM model. The FEM solver takes about 2 sec for every new model simulation. When, the EESM design variables are manipulated to successfully maximize the SFF considering the constraints and search bounds, the design optimization process is ended.

### III. Results and Discussion

#### A. Design Optimization Using PIBO-MESA

The Bayesian-based design optimization process drives iteratively to fit a probabilistic surrogate model to stated values of a cost function. Therefore, a physics-informed feature can administer the future queries to explore the best areas within the search space to: (i) overcome the uncertainty of the surrogate model, and (ii) minimize the cost function much faster than other population-based methods. The developed PIBO-MESA uses probabilistic ML methods to offer a promising framework for understanding the uncertainty of the design problem and its performance predictions. The BO algorithm's uncertainty estimation is calibrated based on the calibrator in [41], in which the calibrated approach found the global minimum prior to the uncalibrated method, as shown in Fig. 5, for the cost function. In Fig. 6(a), the regret minimization using PIBO-MESA for these surrogate models is demonstrated. The GP method has shown the best performance on average. The physics-informed design process administers the superior search regions, as shown in Fig. 6(b). It shows how the initial 4D filled search regions are shrunk iteratively from green to red, and purple at the end of the optimization. For the design optimization problem defined in section II, the performance of the most popular surrogate models is given in Table IV. In this table, the calibration outcomes of the PIBO-MESA using different ML surrogate models is reported. Considering the regret function, the best performance is reported using GPs and NSGA-II, whereas the GP model is 45.85% faster than the NSGA-II method. The reported values are averaged by 20 different random initialization processes. The calibration error is ignored while regression is calculated for the recalibrated models, recalibration only is performed during BO. The regret on the objective function is given in the last iteration of the PIBO. Fig. 7 indicates the copper SFF for the presented trapezoidal stator slot. AC winding loss can be produced in the copper when the injected current is alternating because of both skin and proximity effects. These loss mechanisms are especially high at high-speed operation, when the AC winding loss can become much higher than the DC winding loss. The typical two layers, four identical type 8 Litz wires benchmark model with 60% SFF is shown in Fig. 7(a). The other two slots are optimized SFF with different settings using keystone shaped Litz 8 wires for the best packing density. In Fig. 7(b), the maximum SFF of 80% is achieved using keystone shaped wires with two layers bunched Litz wire 8. Due to the benefits of more winding layers to reduce the AC and DC winding losses, a tradeoff should be done to select the best number of layers and SFF. The second optimized solution, as shown in Fig. 7(c), is set to four layers; this yields a SFF of 75% but with more flexibility for different winding configurations and lower AC winding loss. The number of parallel paths is set to two, where the impedance for each parallel path is considered to be the same to prevent current unbalance and additional copper loss. The winding copper losses are examined with sinusoidal current excitations.

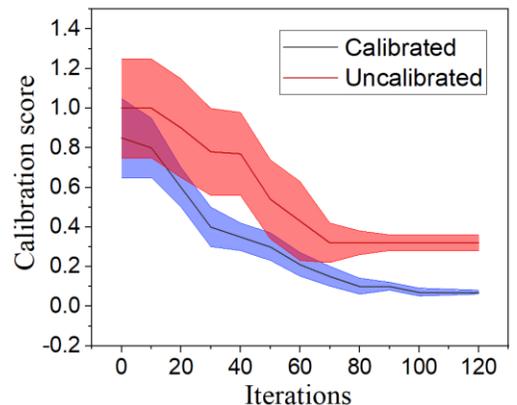

Fig. 5. Calibrating the PIBO-MESA using GP.

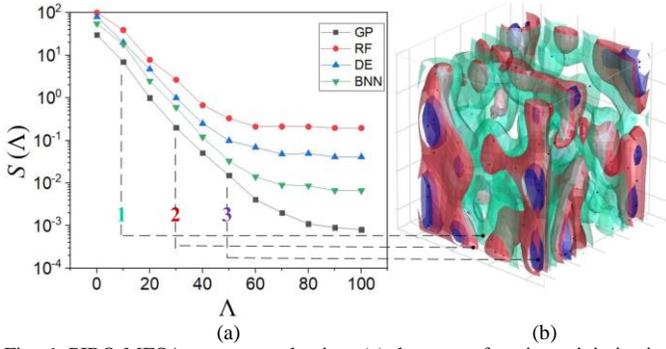

Fig. 6. PIBO-MESA process evaluation, (a) the regret function minimization using different methods, and (b) shrinking iterative 4D search space.

TABLE IV  PERFORMANCE OF SURROGATE MODELS FOR PIBO-MESA

| Average metrics | $S(\Lambda)$ | LSE | CLE | CPUT |
|---|---|---|---|---|
| PIBO-MESA Methods | | | | |
| RF | 0.1965 | 0.6535 | 0.4473 | 5850.45 |
| DE | 0.0455 | 0.7732 | 0.3267 | 5692.82 |
| BNN | 0.0067 | 0.4328 | 0.6653 | 5430.32 |
| GP | **0.0008** | **0.2102** | **0.1997** | **5403.32** |
| Popular Mimic-based Methods | | | | |
| NSGA-II | 0.0009 | 0.2213 | - | 9977.55 |
| PSO | 0.0012 | 0.2466 | - | 9853.29 |

**Note** LSE, CLE, and CPUT are least square error, classification error, and CPU time (in seconds) to complete the design optimization process. Bold and underlined values show the best and worst cases.

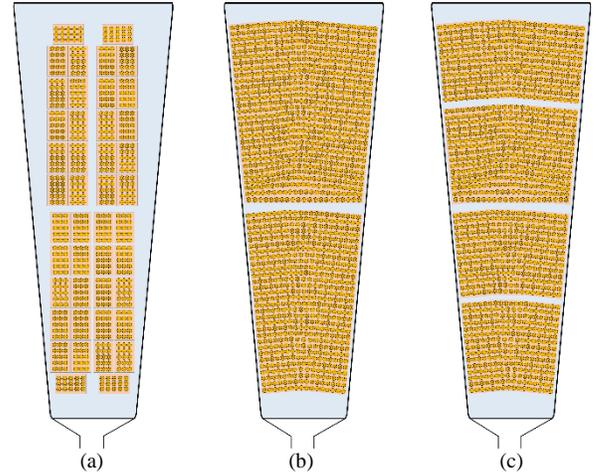

Fig. 7. Slot filling using different methods, (a) benchmark design with four identical type 8 Litz wires, (b) two-layer type 8 Litz, keystone shaped wires winding using the PIBO-MESA with 80% SFF, (c) four-layer type 8 Litz keystone shaped wires winding using the PIBO-MESA with 75% SFF.

*B. EESM New Design and Validation Using FEM*

The electromagnetic results are presented to verify the outcome of the design optimization using PIBO-MESA, in which GP is chosen as the surrogate model. Fig. 8 demonstrates the no-load magnetic flux density distribution of the models. The simulations are done under 200 A and 6000 rpm. For all three models, the maximum magnetic field intensity is observed at the bottom of the stator teeth near the slot opening. In Fig. 8(a), the model 1 (M1) with two-layer (L = 2), four identical, type 8 Litz wires and SFF of 60% shows a maximum magnetic intensity of 1.75 T. Using similar number of layers, the SFF is improved by 20% in the second model (M2), as presented in Fig. 8(b), where 1.88 T is seen in the same spots. To reduce the AC copper loss and efficiency capabilities at high speeds, a greater number of winding layers (L = 4) is used, however that reduced the SFF by 5%. The third model's maximum magnetic intensity is 1.81 T at the same spots.

Table V presents the AC and DC losses for the studied EESM motors. The results show that both optimized solutions (M2 and M3) provided a lower copper loss and higher efficiency during both low and high speed operations. Note that the only difference between M2 (with 80% SFF) and M3 (with 75% SFF) is the number of winding layers, in which M2 and M3 have two and four layers, respectively. The M2 motor can achieve a better performance during the low speeds, whereas the M3 motor provided lower AC copper loss at high speed operations. After a trade-off between AC and DC winding losses, the optimized M2 motor is shown the best performance.

Fig. 9 illustrates the magnetic iron loss density, i.e. hysteresis and Eddy-current losses, influenced by the new designs with higher SFF. All models show the highest iron losses at the bottom of the stator teeth and rotor outer surface close to the stator core. Figs. 9(a1-c1) present the hysteresis loss density, in which the maximum hysteresis loss densities of M1, M2, and M3 motors are 435 kW/m$^3$, 571 kW/m$^3$, and 513 kW/m$^3$, respectively. The Eddy-current loss densities, as presented in Figs. 9(a2-c2), in which the maximum Eddy-current loss densities are 711 kW/m$^3$, 1195 kW/m$^3$, and 1058 kW/m$^3$, respectively.

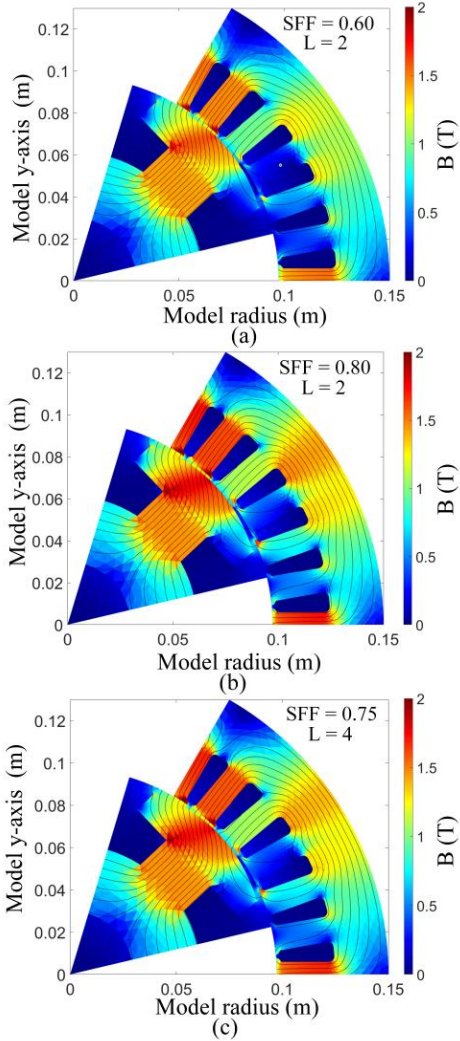

Fig. 8. No load magnetic flux density distribution, (a) M1: benchmark motor, (b) M2: two-layer keystone shaped wound motor using the PIBO-MESA with 80% SFF, (c) M3: four-layer keystone shaped wound motor using the PIBO-MESA with 75% SFF.

TABLE V  EESM'S AC AND DC COPPER LOSSES

| Models/ Parms. | M1 | M2 | M3 |
|---|---|---|---|
| $n$ = 1000 rpm, 200A | | | |
| AC loss (kW) | 0.041 | 0.038 | **0.026** |
| DC loss (kW) | 0.347 | **0.331** | 0.352 |
| Total loss (kW) | 0.388 | **0.369** | 0.378 |
| Improved by (%) | - | **4.897** | -2.577 |
| $n$ = 12,000 rpm, 200A | | | |
| AC loss (kW) | 0.909 | 0.882 | **0.795** |
| DC loss (kW) | 0.408 | **0.405** | 0.416 |
| Total loss (kW) | 1.317 | 1.287 | **1.211** |
| Improved by (%) | - | 2.278 | **8.049** |

**Note** that bold values show the best cases for every parameter.

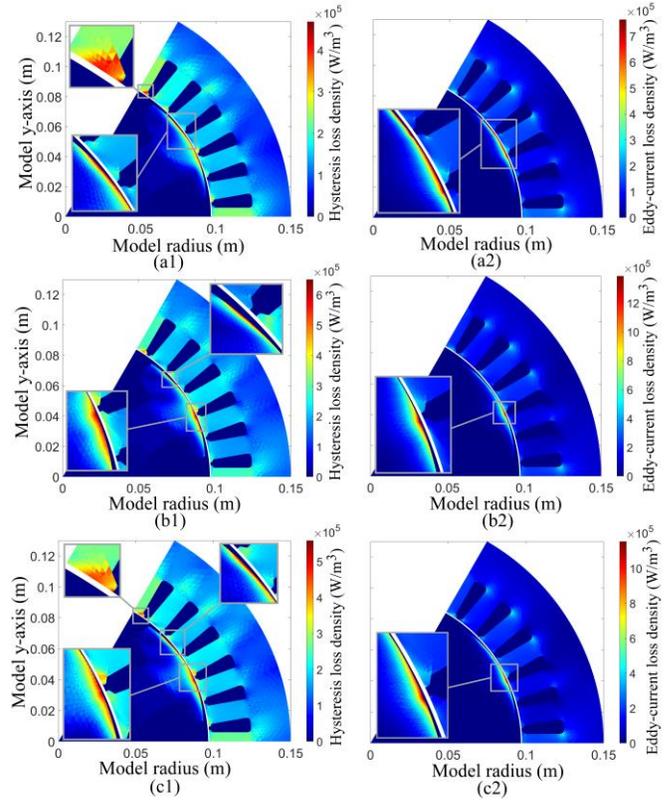

Fig. 9. No load iron loss density distribution, (a) M1: benchmark motor, (b) M2: two-layer wound motor using the PIBO-MESA with 80% SFF, (c) M3: four-layer wound motor using the PIBO-MESA with 75% SFF.

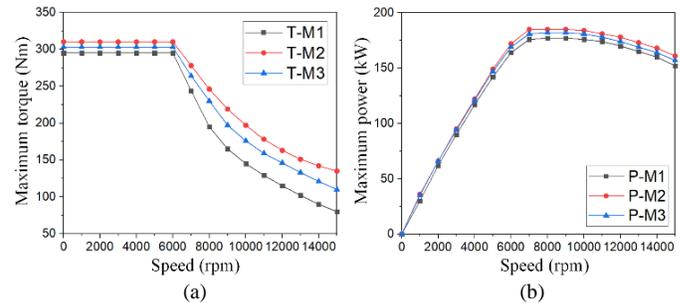

Fig. 10. Comparative peak torque and power profiles, (a) maximum torque, and (b) the maximum power in a wide speed range.

Fig. 10 demonstrates the peak torque and power for the studied models. The results are obtained under a DC-link voltage of 400 V. The maximum electromagnetic torque of M1 with 280 Nm is improved due to higher SFF. As presented in Fig. 10(a), the optimized M2 and M3 motors enhanced the maximum torque by 11.78% and 9.28%, respectively. In Fig. 10(b), the maximum power of M1 motor with 175 kW is also slightly increased, with the M2 and M3 motors offering 1.14% and 2.28% power increase, respectively.

Fig. 11 evinces the efficiency maps of all the studied motors under similar conditions for a fair comparison. The efficiency maps are produced at a 400 V DC-link voltage without consideration of inverters power loss. Fig. 11(a) shows the efficiency map of the benchmark motor, in which the highest efficiency is 93.6% achieved at low torque and medium speed (acceleration phase). The M2 motor, as shown in Fig. 11(b), achieves a higher efficiency (94.9%) and a larger premium efficiency region. The efficiency is enhanced by approximately 1.39% in the premium efficiency region. Fig. 11(c) indicates the efficiency map of the M3 motor with a peak efficiency of 94.2%. The M3 motor increased the efficiency by about 0.64% compared to M1. The efficiency is improved the most in M2 across a wide range speed. Whereas the M3 sacrificed some SFF in order to have lower AC losses, but as presented in Table V this is cancelled by higher DC losses.

Increasing SFF eases the amount of air inside the slot which is being replaced by copper. Air is a weak thermal conductor,

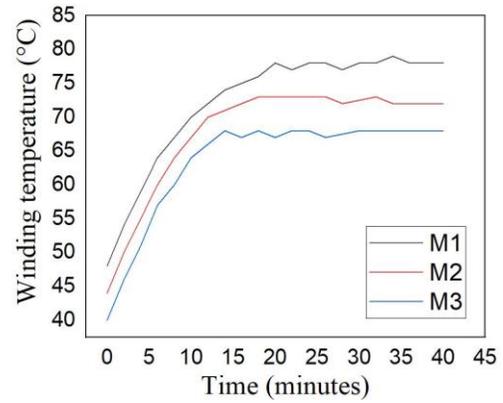
Fig. 12. Transient winding temperature for the studied EESMs.

hence, a decrease in the amount of air in the slot might substantially improve the thermal conduction from the conductors to the cooling liquid ducts. The improved EESMs (M2 and M3) are within the allowed temperature rise, as presented in Fig. 12. The slot area (including winding) temperature computed using the same method as in [44], the results show the case where a peak phase current of 240 A is injected in the stator windings. The steady-state temperature of M2 and M3 motors is decreased by 13.4% and 9.5%, respectively.

IV. CONCLUSIONS

The electromagnetic contribution of the stator copper winding is vital in most traction electrical machines, such as permanent magnet synchronous and induction machines, used in electric vehicles. One of the promising traction electrical machines for low-cost and high-performance applications is magnet-free brushless electrically-excited synchronous machines (EESM). In this work, a new physics-informed Bayesian design optimization (PIBO) method for improving the slot filling factor (SFF) in such machines is presented. In this design optimization, a maximum entropy sampling algorithm (MESA) is used to seed an iterative PIBO algorithm, where the target function and its approximations are produced by Gaussian processes (GPs). The proposed PIBO-MESA worked exclusively with a FEM solver to perform the GP surrogate to achieve an optimal combination of design variables. Significant computational gains were achieved using the new PIBO-MESA approach, which was 45% faster than existing stochastic methods, such as the non-dominated sorting genetic algorithm II (NSGA-II). The proposed PIBO-MESA was coupled with a 2D finite element model (FEM) to perform a GP-based surrogate and provide the first demonstration of the optimal combination of design variables. The FEM results confirm that the new design optimization process tends to a new approach to stator slot winding filling using keystone shaped wires, leading to a higher SFF (i.e. by 20%) and electromagnetic improvements (e.g. maximum torque by 12%) with similar resistivity. The proposed motor achieved an SFF of 80%, and its maximum power is also increased by 2.28%. In absolute terms, the PIBO-MESA design optimization process presented here represents a significant pathway in the faster design and production of future high-performance electrical machines.

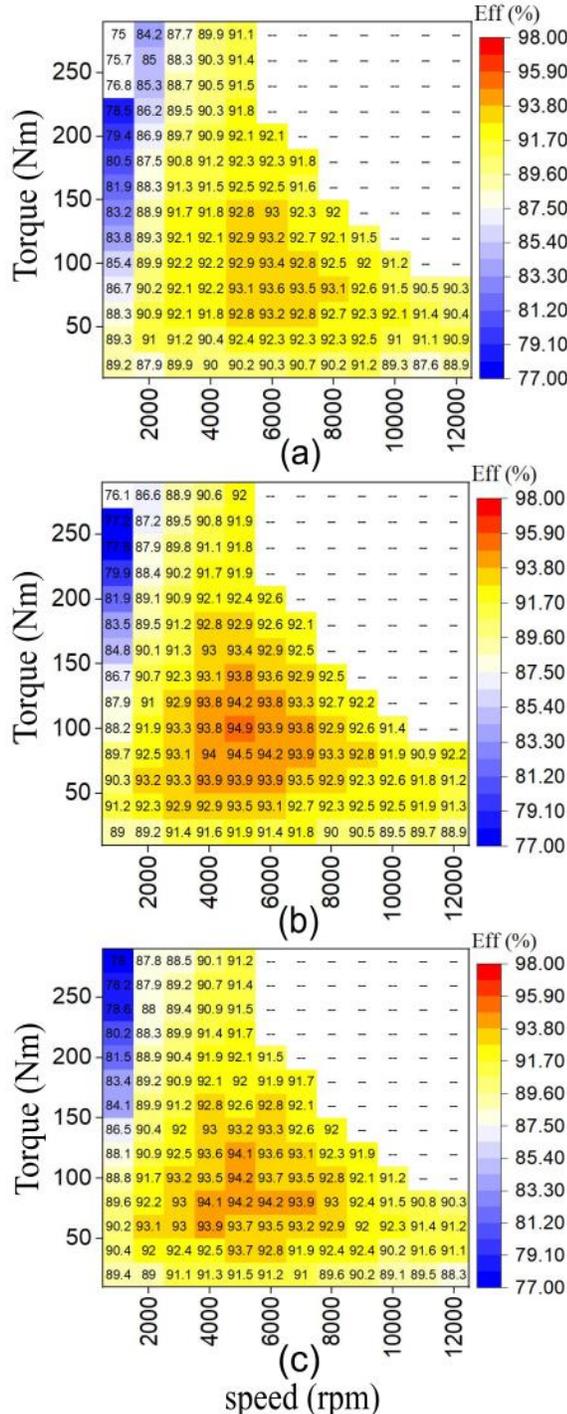
Fig. 11. Efficiency maps for the studied EESMs, M1: benchmark motor, (b) M2: two-layer, keystone shaped, wound motor using the PIBO-MESA with 80% SFF, (c) M3: four-layer, keystone shaped, wound motor using the PIBO-MESA with 75% SFF.

**Data availability statement**:

The generated data is partially available upon request from the authors. Data requests can be made via this email: pedram.asef@ucl.ac.uk.

**Author contributions:**

Conceptualization: [Pedram Asef, Christopher Vagg], ; Methodology: [Pedram Asef]; Formal analysis and investigation: [Pedram Asef, Christopher Vagg]; Writing - original draft preparation: [Pedram Asef, Christopher Vagg]; Writing - review and editing: [Pedram Asef, Christopher Vagg]; Funding acquisition: [Pedram Asef, Christopher Vagg]; Resources: [Pedram Asef, Christopher Vagg]; Supervision: [Pedram Asef, Christopher Vagg].